\documentclass[twocolumn]{aastex7}
\usepackage{comment}

\newcommand\blackholes{BHs }
\newcommand\blackhole{BH }

\newcommand{\f}[2]{\frac{#1}{#2}}
\newcommand{\mr}[1]{\mathrm{#1}}
\newcommand{\p}{\partial}
\newcommand{\h}{\hspace{1mm}}
\newcommand{\bra}[1]{\left(#1\right)}
\newcommand{\ctext}[1]{\raise0.2ex\hbox{\textcircled{\scriptsize{#1}}}}
\newcommand{\chH}{\mathrm{H}}
\newcommand{\chHp}{\mathrm{H}^{+}}
\newcommand{\chHm}{\mathrm{H}^{-}}
\newcommand{\chHt}{\mathrm{H}_{2}}
\newcommand{\chHe}{\mathrm{He}}
\newcommand{\chHep}{\mathrm{He}^+}
\newcommand{\che}{\mathrm{e}^{-}}
\newcommand{\chD}{\mathrm{D}}
\newcommand{\chDp}{\mathrm{D}^{+}}
\newcommand{\chDm}{\mathrm{D}^{-}}
\newcommand{\chHD}{\mathrm{HD}}
\newcommand{\chHDp}{\mathrm{HD}^+}

\newcommand{\eqref}[1]{(\ref{#1})}

\begin{document}

\title{Massive Black Hole Seed Formation in Strong X-ray Environments at High Redshift}

\author[orcid=0000-0001-8382-3966]{Kazutaka Kimura}
\affiliation{Astronomical Institute, Graduate School of Science, Tohoku University, Aoba, Sendai 980-8578, Japan}
\email[show]{kimura.k@astr.tohoku.ac.jp}  

\author[orcid=0000-0001-9840-4959]{Kohei Inayoshi} 
\affiliation{Kavli Institute for Astronomy and Astrophysics, Peking University, Beijing 100871, China}
\email{inayoshi@pku.edu.cn}

\author[orcid=0000-0001-5922-180X]{Kazuyuki Omukai}
\affiliation{Astronomical Institute, Graduate School of Science, Tohoku University, Aoba, Sendai 980-8578, Japan}
\email{fakeemail3@google.com}

\begin{abstract}
Direct collapse of pristine gas in early galaxies is a promissing pathway for forming supermassive black holes (BHs) powering active galactic nuclei (AGNs) at the epoch of reionization (EoR).
This seeding mechanism requires suppression of molecular hydrogen (H$_2$) cooling during primordial star formation via intense far-ultraviolet radiation from nearby starburst galaxies clustered in overdense regions.
However, non-detection of 21~cm signals from the EoR reported by the Hydrogen Epoch of Reionization Array (HERA) experiment suggests 
that such galaxies may also emit X-rays more efficiently than in the local universe, promoting H$_2$ production and thereby potentially quenching massive BH seed formation.
In this study, we examine the thermal and chemical evolution of collapsing gas in dark matter halos using a semi-analytic model incorporating observationally calibrated X-ray intensities.
We find that strong X-ray irradiation, as suggested by HERA, significantly suppresses direct collapse and leads most halos to experience $\chHt$ cooling. 
Nevertheless, massive BH seeds with $M_\mr{BH} \gtrsim 10^4~M_\odot$ still form by $z\simeq 15$, particularly in regions with baryonic streaming motion, 
and their abundance reaches $\sim 10^{-4}~{\rm Mpc}^{-3}$ sufficient to explain the SMBHs identified by JWST spectroscopy at $3<z<6$.
While the formation of highly overmassive BHs with masses comparable to their host galaxies is prohibited by X-ray ionization, our model predicts that BH-to-stellar mass ratios of 
$\simeq 0.01-0.1$ were already established at seeding.
\end{abstract}

\keywords{\uat{Supermassive black holes}{1663} --- \uat{Active galactic nuclei}{16} --- \uat{Quasars}{1319} --- \uat{Black holes}{162} --- \uat{High-redshift galaxies}{734}}


\section{INTRODUCTION}
Supermassive black holes (SMBHs) with masses exceeding $10^9~M_\odot$ have been observed at redshifts $z \gtrsim 6$, corresponding to less than one billion years after the Big Bang \citep[see recent reviews by][]{Inayoshi_et_al_2020,Volonteri_et_al_2021,Fan_et_al_2023}.
The existence of such massive objects at early times poses a significant challenge to our understanding of black hole (BH) formation and growth in the early universe, due to the limited time available for them to reach such masses.
Recent discoveries with the James Webb Space Telescope (JWST) continue to push the detection of SMBHs to even higher redshifts and lower mass regimes, providing new constraints on the formation scenarios 
\citep{Onoue_et_al_2023,Kocevski_et_al_2023,Harikane_et_al_2023_AGN,Maiolino_et_al_2024,Greene_et_al_2024,Matthee_et_al_2024,Kokorev_et_al_2024,Akins_et_al_2024,Juodzbalis_et_al_2024_mnras}.
Intriguingly, a signifiant fraction of these SMBHs appear to be overmassive relative to their host stellar masses, lying significantly above the local $M_\mathrm{BH}$–$M_*$ relation \citep{Kormendy_and_Ho_2013,Reines_and_Volonteri_2015,Pacucci_et_al_2023,Junyao_Li_et_al_2025}.
Massive \blackhole seeds with $M_\mr{BH}\sim10^{4\text{--}6}~M_\odot$ provide a natural explanation for the presence of such SMBHs, and therefore have attracted considerable attention.

\par
Massive BH seeds are considered to form through so-called direct collapse (DC) of massive gas clouds.
In this scenario, suppression of molecular hydrogen ($\chHt$) cooling, which is the primary coolant in primordial gas for driving star formation, is a key requirement.
If $\chHt$ cooling is effectively suppressed, the gas collapses solely via atomic hydrogen cooling while maintaining a high temperature of $\sim8000$~K. Since the accretion rate onto the central object is estimated as $\dot{M}\sim c_\mr{s}^3/G$, where $c_\mr{s}$ is the sound speed and $G$ is the gravitational constant \citep[e.g.,][]{Shu_1977}, this collapse enables high accretion rates of $\sim0.1\text{--}1~M_\odot~\mr{yr}^{-1}$, leading to the formation of a supermassive star with $\gtrsim 10^4~M_\odot$ \citep{Bromm_and_Loeb_2003, Hosokawa_et_al_2012_SMS,Inayoshi_and_Omukai_2014,Becerra_et_al_2015,Toyouchi_et_al_2023}. 
The supermassive star eventually collapses into a BH of comparable mass due to general relativistic instability 
\citep{Shibata_and_Shapiro_2002, Umeda_et_al_2016, Woods_et_al_2017}.
In contrast, if $\chHt$ cooling is not suppressed, the gas cools down to $\sim 200$~K and undergoes fragmentation, resulting in the formation of typical Population III stars with masses of $\ll 10^{3}~M_\odot$ \citep{Hirano_et_al_2014, Hirano_et_al_2015, Susa_et_al_2014, Sugimura_et_al_2020, Sugimura2023}. 
\par

The most widely considered mechanism for
suppressing $\chHt$ cooling is $\chHt$-photodissoiating Lyman-Werner (LW) radiation. 
The required LW intensity $J_\mr{crit}$ is as high as $J_\mr{crit,21}\simeq 10^{2\text{--}5}$ in the units of $10^{-21}~\mr{erg~s}^{-1}~\mr{cm}^{-2}~\mr{Hz}^{-1}~\mr{sr}^{-1}$, 
depending on the spectral shape of the radiation sources \citep{Omukai_2001,Shang_et_al_2010,Inayoshi_and_Omukai_2011,Sugimura_et_al_2014,Sugimura2016, Latif_et_al_2014,Agarwal2016}.
This critical value is substantially higher than the typical background LW intensity in the local universe ($J_{21}\sim10^{-3}$--$10^{-2}$) and
the expected value at high redshifts ($J_{21}\sim10^{-1}$--$10^1$) \citep{Haardt_and_Madau_2012,Johnson_et_al_2013}.
Therefore, one must consider locally enhanced LW intensity resulting from halo clustering.
Even when halo clustering is accounted for, however, only an extremely small fraction of halos are exposed to LW fluxes exceeding $J_\mr{crit}$ \citep{Dijkstra_et_al_2008,Dijkstra_et_al_2014}.
Strong halo clustering in overdense regions is thus particularly important to achieve such high LW intensities \citep{Li_et_al_2021,Lupi_et_al_2021}.
Alternatively, recent studies suggest that, even if the LW intensity is below $J_\mathrm{crit}$, baryonic streaming motion 
\citep{Hirano_et_al_2017,Schauer_et_al_2017} and dynamical heating by frequent mergers \citep{Wise_et_al_2019,Sakurai_et_al_2020} 
may play a critical role in suppressing $\mathrm{H}_2$ formation and enabling direct collapse black hole (DCBH) formation \citep{Inayoshi_et_al_2018}.
\par

While LW irradiation by nearby star-forming galaxies is one of the main drivers for DC, these galaxies should also produce high-energy radiation and relativistic particles, possibly originating from high-mass X-ray binaries, micro-quasars, and energetic supernova explosions \citep{Stacy2007, Inayoshi_and_Omukai_2011,Inayoshi_and_Tanaka_2015}. 
Both X-rays and cosmic-rays increase the ionization degree in the gas, promote the production of $\chHt$ through the electron-catalyzed reaction, and potentially hinder DC.  
Including the X-ray ionization, the critical LW flux $J_\mr{crit}$ substantially increases as $J_\mr{crit} \propto J_{\rm X}^{1/2}$, where $J_{\rm X}$ is the X-ray intensity from the same radiation sources \citep{Inayoshi_and_Omukai_2011}.
Given the observed local relation between X-ray luminosity and star formation rate (SFR), this scaling implies a significant increase in $J_\mr{crit}$, making the formation of DCBHs extremely difficult.
Recent non-detection of 21 cm power spectrum by Hydrogen Epoch of Reionization Array (HERA) further suggests that the X-ray production efficiency in the early universe could be $10\text{--}240$ times higher than in the local values \citep{Abdurashidova_et_al_2022,HERA_Collaboration_2023,Lazare_et_al_2024}.
This implies that the suppression of DCBH formation by X-ray irradiation could be even more severe.

\par
In this work, we investigate seed formation in rare, overdense regions by incorporating the X-ray effects, in order to provide a more realistic black hole mass function (BHMF).
Previous studies have typically focused on the regions with $\sim2\sigma$ mass variance, where achieving the critical LW flux is extremely rare \cite[e.g.][]{Dijkstra_et_al_2008,Dijkstra_et_al_2014}.
In contrast, high‑redshift quasars are considered to reside in even rarer environments $\gtrsim 3\sigma$, 
characterized by significantly elevated halo clustering. 
Recent theoretical studies suggest that the increased efficiency of DCBH formation in these clustered regions can compensate for their overall scarcity in cosmic volume \citep{Lupi_et_al_2021,Li_et_al_2021,Li_et_al_2023}.
To study BH seed formation in such regions, we extend the semi-analytic models of \citet{Li_et_al_2021} by including X-ray radiation from nearby galaxies, with intensities spanning values inferred from both local observations and HERA data.
This allows us to evaluate the impact of elevated X-ray irradiation on the abundance and BHMF in high-$\sigma$ environments. 

\par
While our study was being finalized, a related work by \citet{Zhang_et_al_2025} appeared, which also investigates the impact of X-ray irradiation on DCBH formation. They use a one-zone chemical model to examine how X-ray-induced ionization affects the conditions for DCBH formation, and further explore its implications for the global 21 cm absorption signal from the early universe. Although we also adopt a one-zone model to follow the thermal and chemical evolution of collapsing gas, our study differs in scope: we incorporate this into a semi-analytic framework based on merger trees to evaluate the statistical distribution of BH seed masses under realistic environmental conditions, including baryonic streaming motion. Our work thus provides a complementary perspective, focusing on population-level outcomes rather than local threshold conditions.

\par
This paper is organized as follows. 
In Section \ref{sec:methodology}, we describe the semi-analytic model used to study BH seed formation, including the treatment of X-ray irradiation. 
Section \ref{sec:result} presents the results of our calculations, highlighting the impact of X-ray irradiation on gas thermal evolution and BH seed mass distribution. In Section \ref{sec:discussion}, we discuss the implications of our findings in the context of observed high-redshift SMBH formation and the uncertainty in the effect of baryonic streaming motion.
Finally, we provide a summary in Section \ref{sec:summary}.

\section{METHODOLOGY} \label{sec:methodology}
To study the BH seed formation, we adopt the semi-analytical model developed by \citet{Li_et_al_2021} and implement the effect of X-ray irradiation from nearby galaxies into their model. 
In Section~\ref{subsec:SeedBHForm}, we provide a brief overview of the model, and we refer readers to \citet{Li_et_al_2021} for more detailed descriptions. 
In Section~\ref{subsec:X-ray}, we describe how we incorporate X-ray ionization and heating into the model.

\subsection{BH seed formation} \label{subsec:SeedBHForm}
Following \citet{Li_et_al_2021}, we calculate the evolution of density, temperature, and chemical composition of gas clouds at the center of dark matter (DM) halos that grow via gas accretion and mergers.
In this study, we focus on progenitor halos that end up in a massive halo with $M_\mr{h} \geq 10^{11} M_{\odot}$ at $z=6$, as implied by observations that high-redshift SMBHs typically reside in such massive halos \citep{Shimasaku_and_Izumi_2019}.
Specifically, we consider DM halos with masses of $M_\mr{h} = 10^{11}$, $10^{12}$, and $10^{13} M_{\odot}$ at $z=6$, corresponding to biased overdense regions with $>3\sigma$.
In these overdense regions, frequent halo mergers driven by clustering, along with enhanced radiation from nearby galaxies, establish favorable conditions for massive BH seed formation \citep{Lupi_et_al_2021,Li_et_al_2021}.
Although such clustered regions may already be mildly metal-enriched by prior star formation, recent simulations show that supermassive star formation remains possible under strong far-ultraviolet (FUV) radiation even at $Z \lesssim 10^{-3}\,Z_\odot$ \citep{Chon2020, Chon2024b}.

\par
To quantify the gas supply rate onto DM halos and the frequency of halo mergers, we generate $10^3$ merger trees for each halo mass of interest using {\tt GALFORM}, based on the extended Press-Schechter formalism \citep{Press_and_Schechter_1974,Lacey_and_Cole_1993,Cole_et_al_2000,Parkinson_et_al_2008}. 
In constructing these trees, we set the minimum DM halo mass to $10^5 M_{\odot}$ to properly capture the earliest star formation via $\chHt$ line cooling at high redshifts. 
Heating and turbulence induced by gas accretion and mergers are modeled based on results from numerical simulations \citep{Wise_and_Abel_2007}.
Furthermore, we calculate the time-dependent $\chHt$-photodissociating FUV radiation flux following \citet{Li_et_al_2021}, assuming Pop II star formation with a star formation efficiency of 0.05.
Additionally, we assume that both X-ray and FUV fluxes scale with the SFRs of source galaxies, and in turn, the fluxes are linearly correlated.
The value of the scaling factor is treated as a free parameter within the range suggested by local observations and HERA measurements. 
We provide a detailed explanation in Section \ref{subsubsec:X-ray_flux}.
Our formulation does not account for halo-to-halo scatter in the FUV or X-ray flux intensities.
However, since strong radiation fields in our model typically arise from the combined contributions of several tens of galaxies in our model, we expect that such scatter has a negligible impact on the results.
\par
We also consider the baryonic streaming motion relative to DM, which originates from the epoch of cosmic recombination at $z_\mathrm{rec} \simeq 1100$. 
The baryonic streaming motion delays the infall of gas into the DM potential and maintains a high level of kinetic energy in the gas, thereby suppressing gravitational collapse.
We perform calculations for two cases: $v_\mathrm{bsm} = 0$ and $v_\mathrm{bsm} = 1\sigma_\mathrm{bsm}$, where the root-mean-square (rms) speed of baryonic streaming motion at a given redshift $z$ is given by $\sigma_\mathrm{bsm} = 30~\mathrm{km}~\mathrm{s}^{-1}~  (1+z)/(1+z_\mathrm{rec})$.
Using this, we estimate the effective sound speed of the gas at the center of the halo as 
\begin{equation}
c_\mathrm{eff}^2 = {c_\mr{s}^2 + \frac{v_\mathrm{tur}^2}{3} + (\alpha_0 v_\mathrm{bsm})^2}, \label{eq:c_eff}
\end{equation}
where $c_s$ is the thermal sound speed, $v_\mathrm{tur}$ is the turbulent velocity generated by gas accretion and halo mergers, and the coefficient $\alpha_0$ is set to $4.7$, the value suggested by \citet{Hirano_et_al_2018}.  
The value of $\alpha_0$ has been investigated in previous studies, but remains highly uncertain in the range of $\sim1$--$10$ \citep{Hirano_et_al_2017,Schauer_et_al_2019,Hirano_et_al_2025}. 
We discuss the impact of varying $\alpha_0$ in Section \ref{subsec:bsm}.

\par
Based on the thermal evolution of collapsing gas obtained by the method described above, we evaluate the masses of seed \blackholes formed within halos.
First, we calculate the accretion rate onto the central region of collapsing gas by $\dot{M}=c_\mr{eff}^3/G$ with $c_\mr{eff}$ given by Equation \eqref{eq:c_eff}.
The temperature that determines $c_s$ and the value of $v_\mr{tur}$ are taken at the point during the collapse where the density has exceeded $n_\chH > 10^3~\mr{cm}^{-3}$ and the temperature reaches its minimum.
The accretion rate onto the central protostar is then given by $\dot{M}_*=\eta\dot{M}$, where $\eta$ is the conversion efficiency owing to the accretion via the circumstellar disk.
Based on the results from multi-dimensional simulations \citep{Sakurai_et_al_2016,Toyouchi_et_al_2023}, we adopt $\eta=0.3$.
From this accretion rate, we estimate the final stellar mass with a phenomenological model.
When the accretion rate is below the critical threshold, $\dot{M}_* < \dot{M}_\mr{crit}=0.04~M_\odot~\mr{yr}^{-1}$ \citep{Hosokawa_et_al_2012_SMS}, the protostar undergoes Kelvin–Helmholtz contraction and settles onto the main sequence. 
In this case, its final mass is regulated by photoevaporation feedback and can be approximated as
\begin{eqnarray}
    M_\mr{*,fb} \simeq 2.9 \times 10^3 M_\odot~\bra{\f{\dot{M}_*}{0.01~M_\odot~\mr{yr}^{-1}}} ,
\end{eqnarray}
as derived by \citet{Li_et_al_2021}.
On the other hand, when the accretion rate exceeds the critical value, $\dot{M}_*>\dot{M}_\mr{crit}$, 
the protostar becomes highly inflated, resulting in a low surface temperature. 
This suppresses ultraviolet (UV) emission and renders radiative feedback ineffective.
In this case, the stellar mass is ultimately limited by the onset of the general relativistic instability, given approxomately by \citep{Woods_et_al_2017}:
\begin{eqnarray}
    M_\mr{*,GR} \simeq \left[ 0.83~\log\bra{\f{\dot{M}_*}{M_\odot~\mr{yr}^{-1}}}+ 2.48 \right]\times10^{5}~M_\odot . \nonumber \\
\end{eqnarray}
We assume that mass loss via stellar winds from these massive primordial stars is negligible during their evolution \citep{Inayoshi_et_al_2013, Hosokawa_et_al_2013}, and the remnant seed \blackhole mass $M_\mr{BH}$ is equal to the mass of its progenitor.

\subsection{X-ray implementation} \label{subsec:X-ray}
\subsubsection{Newly added chemical reactions}
X-rays from nearby galaxies heat and ionize the gas of interest. 
X-ray ionization enhances the electron fraction and promotes $\mathrm{H_2}$ formation through electron-catalyzed reactions,
\begin{eqnarray}
    && \chH + \che \rightarrow \chHm + \gamma ,\\
    && \chHm + \chH \rightarrow \chHt + \che,
\end{eqnarray}
and thus lowers the gas temperature via $\chHt$ cooling. 
When the temperature decreases below $\sim 200~{\rm K}$, $\mathrm{HD}$ formation becomes more efficient. 
Therefore, we incorporate $\mathrm{HD}$ chemistry in our chemical reaction network to accurately follow the temperature evolution. The additional reactions are listed in Table \ref{tab:ChemReact} \citep{Inayoshi_and_Omukai_2011}. 
Here, we do not consider the ionization by UV Lyman continuum photons ($h\nu > 13.6~\mr{eV}$) from nearby galaxies.
This is because, unlike FUV and X-ray photons, ionizing photons have large absorption cross sections and cannot escape efficiently from the source galaxies.
Moreover, halos where DCBHs are expected to form tend to be surrounded by dense gas, which further enhances the shielding of ionizing photons.
As shown by cosmological radiation hydrodynamics simulations \citep{Chon_and_Latif_2017}, these effects result in negligible ionizing radiation reaching the DCBH-forming halos.
\begin{table}[t]
    \caption{Newly added chemical reactions.}
    \label{tab:ChemReact}
    \centering
        \begin{tabular}{clc}
        \hline
        Number & Process & Reference \\
        \hline 
         & D reactions &  \\
         \hline
        1 & $ \chDp  + \che \rightarrow \chD + \gamma $ & 1 \\
        2 & $ \chD + \chHp \rightarrow \chDp + \chH $ & 2 \\
        3 & $ \chDp + \chH \rightarrow \chHp + \chD $ & 2 \\
        4 & $ \chD + \chH \rightarrow \chHD + \gamma $ & 1 \\
        5 & $ \chD + \chHt \rightarrow \chH + \chHD $ & 1 \\
        6 & $ \chHDp + \chH \rightarrow \chHp + \chHD $ & 1 \\
        7 & $ \chDp + \chHt \rightarrow \chHp + \chHD $ & 3 \\
        8 & $ \chHD + \chH \rightarrow \chHt + \chD $ & 1 \\
        9 & $ \chHD + \chHp \rightarrow \chHt + \chDp $ & 3 \\
        10 & $ \chD + \chHp \rightarrow \chHDp + \gamma $ & 1 \\
        11 & $ \chDp + \chH \rightarrow \chHDp + \gamma $ & 1 \\
        12 & $ \chHDp + \che \rightarrow \chH + \chD $ & 1 \\
        13 & $ \chD + \che \rightarrow \chDm + \gamma $ & 1 \\
        14 & $ \chDp + \chDm \rightarrow 2\chD $ & 1 \\
        15 & $ \chHp + \chDm \rightarrow \chD + \chH $ & 1 \\
        16 & $ \chHm + \chD \rightarrow \chH + \chDm $ & 1 \\
        17 & $ \chDm + \chH \rightarrow \chD + \chHm $ & 1 \\
        18 & $ \chDm + \chH \rightarrow \chHD + \che $ & 1 \\
        19 & $ \chHD + \gamma \rightarrow \chH + \chD $ & 4 \\
        \hline
         & X-ray ionization &  \\
         \hline
        20 & $ \chH + \gamma \rightarrow \chHp + \che $ & 5 \\
        21 & $ \chHe + \gamma \rightarrow \chHep + \che $ & 6 \\
        \hline
        \multicolumn{3}{p{6.47cm}}{References:(1) \citet{Nakamura_and_Umemura_2002}; (2) \citet{Savin_2002}; (3) \citet{Galli_and_Palla_2002}; (4) \citet{Wolcott-Green_and_Haiman_2011}; (5) \citet{Rybicki_and_Lightman_1979}; (6) \citet{Abel_et_al_1997}} 
        \end{tabular}
\end{table}

\subsubsection{X-ray ionization and heating}
In this study, we consider X-rays with photon energies of $2$--$10$~keV. The photoionization cross sections for hydrogen and helium decrease with increasing photon energy, and photons with energies above 2 keV have sufficiently long mean free paths to reach nearby galaxies without being absorbed by the circumgalactic/intergalactic gas.
We assume that the incident X-ray flux has a power-law spectrum with an index of $-1.5$,
\begin{eqnarray}
    && J_\mr{X}(\nu) = J_\mr{X,21} \times 10^{-21} \bra{\f{\nu}{\nu_0}}^{-1.5} \nonumber \\ 
    && \hspace{2cm} \mr{erg}~\mr{s}^{-1}~\mr{cm}^{-2}~\mr{sr}^{-1}~\mr{Hz}^{-1} , \label{eq:X-ray_Spec}
\end{eqnarray}
where $h\nu_0=1$~keV. Here, $J_\mr{X,21}$ denotes the normalization of the X-ray flux with its prescription given in Section \ref{subsubsec:X-ray_flux}.
\par
We evaluate the ionization and heating rates following \citet{Inayoshi_and_Omukai_2011}. While we provide a brief overview here, we refer readers to the original paper for further details.
We estimate the primary ionization rate for each species ($\chH$ and $\chHe$) using the X-ray spectrum and the optical depth.
Electrons produced by primary ionization have energies significantly larger than the ionization potentials, leading to secondary ionization of hydrogen and helium. 
The total ionization rate is given by the sum of the primary and secondary rates, which we use as the reaction rates for reactions 20 and 21 in Table \ref{tab:ChemReact}.
The heating rate due to X-ray ionization is determined by the energy deposited as heat per ionization event. 
We estimate the secondary ionization rate and the heating rate using fitting formulae provided by \citet{Wolfire_et_al_1995}.

\subsection{X-ray intensity} \label{subsubsec:X-ray_flux}
The value of $J_{\mr{X},21}$ in Equation (\ref{eq:X-ray_Spec}) determines the X-ray intensity received by halos potentially hosting BH formation. 
We assume that $J_{\mr{X},21}$ is proportional to dimensionless LW intensity $J_{21}$ and treat its proportionality constant as a free parameter.
In our model, $J_{21}$, which is defined as $J_\mathrm{FUV}(\nu_L)/(10^{-21}~\mr{erg}~\mr{s}^{-1}~\mr{cm}^{-2}~\mr{Hz}^{-1}~\mr{sr}^{-1})$, where $h\nu_L = 13.6~\mathrm{eV}$, represents the intensity of LW radiation from nearby galaxies.
This intensity evolves as the host halo grows over time.
This approach is motivated by the fact that both X-rays and LW radiation trace massive star-forming activity, and therefore both are expected to scale with the SFR. 
Although X-ray sources are generally believed to be high-mass X-ray binaries, there remain large theoretical uncertainties in modeling their formation efficiency and X-ray luminosity, particularly in the high-redshift universe \citep{Mirabel_et_al_2011,Fragos_et_al_2013}. 
Therefore, we explore a parameter range of the proportionality constant based on the observations.
\par
First, we estimate the proportionality constant from local observations.
Local star-forming galaxies exhibit a correlation between the 2--10~keV X-ray luminosity and the SFR, typically expressed as $L_{\mathrm{X},2\text{--}10~\mr{keV}}/\mr{SFR} \sim 6 \times 10^{39}~\mathrm{erg~s^{-1}~M_\odot^{-1}~yr}$ \citep[][]{Glover_and_Brand_2003}.
Combining this value with the spectral shape of Equation \eqref{eq:X-ray_Spec}, we can get the value of $J_\mr{X,21}$ as
\begin{eqnarray}
    && J_\mr{X,21} = 4.2\times10^{-3} \nonumber \\
    && \hspace{1cm} \bra{\f{d}{10~\mr{kpc}}}^{-2}~\bra{\f{\mr{SFR}}{20~M_\odot~\mr{yr}^{-1}}} , \label{eq:JX21_local}
\end{eqnarray}
where $d$ is the distance from the source. 
Meanwhile, we estimate the LW intensity based on theoretical considerations.
We consider a star-forming galaxy with a metallicity of $10^{-3}~Z_\odot$ and a Salpeter initial mass function ranging from 1 to 100~$M_\odot$, whose radiation is approximated by a blackbody spectrum with a temperature of $2 \times 10^4$~K \citep{Inoue_2011,Sugimura_et_al_2014}.
Using the LW photon emissivity with energies between 11.2 eV and 13.6 eV from \citet{Schaerer_2003}, we can estimate $J_{21}$ as
\begin{eqnarray}
    J_{21} = 1.5 \times 10^3 \left(\frac{d}{10~\mathrm{kpc}}\right)^{-2} \left(\frac{\mathrm{SFR}}{20~M_\odot~\mathrm{yr}^{-1}}\right). \label{eq:J21}
\end{eqnarray}
From Equations \eqref{eq:JX21_local} and \eqref{eq:J21}, the ratio of X-ray to LW intensity $J_\mr{X,21}/J_{21}$ is $\sim2.8\times10^{-6}$. 
\par
In contrast, the non-detection of the 21 cm power spectrum at redshifts $z = 7.9$ and $z = 10.4$ by HERA indicates a higher level of X-ray intensities.
This is because X-ray heating raises the intergalactic medium spin temperature, reduces the 21 cm signal contrast against the cosmic microwave background, and thereby suppresses the power spectrum.
A Bayesian analysis using the semi-analytic model 21cmFAST suggests that the X-ray luminosity per SFR is about 10--240 times higher than the local value \citep{Abdurashidova_et_al_2022,HERA_Collaboration_2023,Lazare_et_al_2024}. 
Theoretical studies are consitent with this scenario, predicting enhanced X-ray luminosity in the early universe due to factors such as low metallicity, a top-heavy initial mass function, and efficient formation of high-mass X-ray binaries \citep{Fragos_et_al_2013}.
Therefore, in this work, 
we perform calculations with $J_\mr{X,21}/J_{21}$ set to $0$, $10^{-6}$, $10^{-5}$, and $10^{-4}$ to explore the impact of varying X-ray intensities.

\section{RESULTS} \label{sec:result}

\subsection{The impact of varying X-ray intensity}

We first examine how the collapse of gas clouds and the resulting BH seed formation depend on the intensity of X-ray irradiation.
Here, we focus on the cases where the host DM halos grow to $M_{\rm h}=10^{12}~M_\odot$ at $z=6$.
The overall trend of thermal evolution in these massive halos is qualitatively similar in halos with $M_{\rm h}=10^{11}$ and $10^{13}~M_\odot$.
In Section~\ref{sec:BHMF}, we combine the results for all DM halo masses to assess the statistical properties of the entire BH population.
\par
\begin{figure*}[t]
    \begin{center}
      \includegraphics[width=\linewidth]{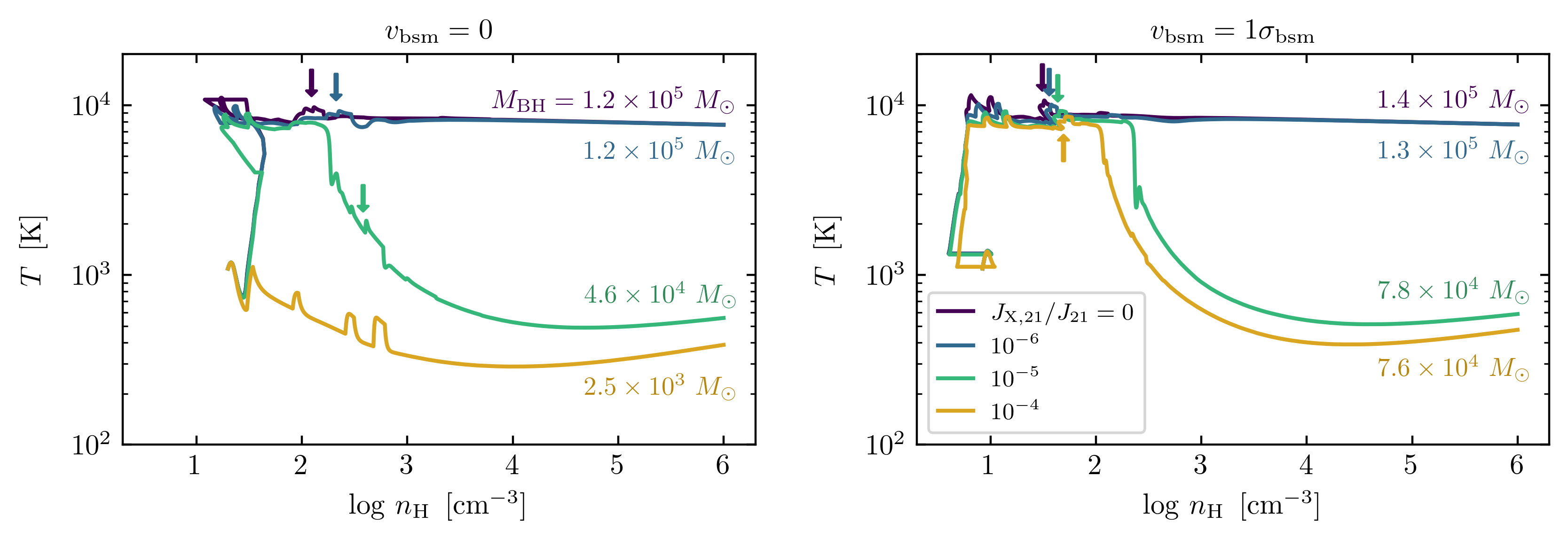}
      \caption{
      Gas thermal evolution in the main progenitor of a DM halo that grows to $10^{12}~M_\odot$ at $z=6$, computed for $v_\mathrm{bsm}=0$ (left panel) and $1\sigma_\mathrm{bsm}$ (right panel). 
    The line colors represent different X-ray intensities: $J_\mathrm{X,21}/J_{21} = 0$ (purple), $10^{-6}$ (blue), $10^{-5}$ (green), and $10^{-4}$ (dark orange). 
    Colored arrows indicate the moments when $J_{21}$ reaches $10^3$ along each corresponding track. 
    In the case of $v_\mathrm{bsm} = 0$ and $J_\mathrm{X,21}/J_{21} = 10^{-4}$, the gas collapses before $J_{21}$ reaches $10^3$ due to X-ray-induced H$_2$ formation and cooling, and thus no arrow is shown. 
    The resulting BH seed mass for each track is labeled in the corresponding color directly in the figure.
    This figure demonstrates how X-ray irradiation and baryonic streaming motion affect gas thermal evolution and BH seed formation. 
    In particular, strong X-ray backgrounds can trigger early H$_2$ formation, altering the collapse pathway, while baryonic streaming delays collapse and suppresses the H$_2$-cooling track.
        }
      \label{fig:EvolOnezone}
    \end{center}
\end{figure*}

Figure~\ref{fig:EvolOnezone} presents the temperature evolution of a collapsing cloud in a DM halo irradiated by both LW radiation and X-rays from nearby galaxies.
The left and right panels correspond to cases without baryonic streaming motion ($v_\mr{bsm}=0$) and with a typical streaming motion ($v_\mr{bsm}=1\sigma_\mr{bsm}$), respectively.
To isolate the effects of X-rays, we compare four cases with different X-ray intensities:
$J_\mr{X,21}/J_\mr{21}=0$ (purple), $10^{-6}$ (blue), $10^{-5}$ (green), and $10^{-4}$ (dark-orange).
The thermal evolution is tracked along a single merger tree of the main progenitor of DM halos that eventually grow to $M_{\rm h}=10^{12}~M_\odot$ at $z=6$.
Along this tree, the LW intensity builds up to $J_{21}\gtrsim 10^{3}$ due to halo clustering, well above the critical LW intensity for enabling DCBH formation.
The colored arrows mark the points along each track where $J_{21}$ reaches $10^{3}$.
We also label the mass of the resulting BH seed for each track in the corresponding color in the figure.
\par
Before discussing the effect of varying X-ray intensities, we comment on the fluctuations observed in all tracks at densities below $10^3~\mathrm{cm}^{-3}$.
These features result from dynamical heating associated with major mergers. 
In our model, before the gas becomes gravitationally unstable due to efficient cooling, the density at the halo center is determined by solving a hydrostatic equation. 
If a major merger occurs during this phase, the energy injection via dynamical heating, leading to a new hydrostatic configuration with reduced central density.
Once cooling becomes efficient, gravitational collapse begins, and the density increases on the free-fall timescale. 
During this collapse phase, subsequent mergers can still inject thermal energy and induce sharp rises in temperature. 
However, once the density exceeds $10^3~\mathrm{cm}^{-3}$, the collapse proceeds so rapidly that there is little time for mergers to affect the thermal evolution.

\par

When the X-ray flux is weak ($J_\mr{X,21}/J_\mr{21}\leq10^{-6}$), the gas remains at a high temperature of $T\simeq 8000~\mr{K}$ during the collapse phase due to H$_2$-photodissociating radiation, regardless of the streaming motion.
In the absence of H$_2$ cooling, the cloud collapse proceeds via atomic hydrogen cooling through Ly$\alpha$ and H$\alpha$ bound-bound transitions,
as well as bound-free and free-free transitions of H$^-$ ions, maintaining the gas temperature at 
$T\sim 5000\text{--}8000~\mr{K}$ \citep{Omukai_2001}.
While the figure shows the evolution up to $n_{\rm H}=10^6~\mr{cm}^{-3}$, the collapse continues in a runaway fashion 
until the central core becomes opaque to continuum radiation at $n_{\rm H} > 10^{16}~\mr{cm}^{-3}$, i.e., 
the point of protostar formation \citep{Inayoshi_and_Omukai_2014,Becerra_et_al_2015}.
Such collapse at high temperatures ultimately leads to the formation of massive BH seeds with masses of $\sim 10^5~M_\odot$.

\par
As the X-ray flux increases, the electron fraction in the collapsing gas rises due to ionization by X-rays, 
which penetrate deeper into the dense gas as the hydrogen photoionization cross section declines sharply as 
$\nu ^{-3}$ toward $\sim$ keV photon energies.
The enhanced electron fraction promotes H$_2$ formation via electron-catalyzed reactions involving H$^-$.
In the case of $J_\mr{X,21}/J_\mr{21}=10^{-5}$, this X-ray-induced H$_2$ formation becomes significant at 
$n_{\rm H} \sim 10^2~\mathrm{cm}^{-3}$, well before reaching the critical density of $\sim 10^4~\mathrm{cm}^{-3}$ 
where collisional dissociation dominates. 
If sufficient H$_2$ does not form by this point, its cooling remains inefficient, and the gas continues to evolve along the atomic track without transitioning to the H$_2$-cooling track
(the so-called ``zone of no return", 
\citealt{Omukai_2001, Inayoshi_and_Omukai_2012,Fernandez_et_al_2014}). 
However, even in this case, the effective sound speed described by Equation (1) is significantly enhanced due to turbulence triggered by rapid gas accretion and violent mergers in overdense regions.
As a result, the black hole masses reach $4.6\times10^4~M_\odot$ and $7.8\times10^4~M_\odot$ for the cases with $v_\mr{bsm}=0$ and $1\sigma_\mr{bsm}$, which are larger than those of typical Population III stars.
\par
For the highest X-ray intensity ($J_\mathrm{X,21}/J_{21} = 10^{-4}$), the impact is even more pronounced.
In the absence of streaming motion ($v_\mathrm{bsm} = 0$), H$_2$ formation and cooling become effective earlier, allowing the gas to begin collapsing when the LW intensity is still relatively modest ($J_{21} \sim 100$).
The gas remains at a low temperature $\lesssim~10^3~\mr{K}$ during the collapse, which leads to the relatively light BH seed mass of $2.5\times10^{3}~M_\odot$.
In contrast, in the presence of baryonic streaming motion ($v_\mathrm{bsm} = 1\sigma_\mathrm{bsm}$), the collapse is delayed until a later time when the background LW intensity is significantly higher (see also Figure~\ref{fig:RedshiftDistribution}).
Although the gas temperature initially rises to $\sim 10^4~\mr{K}$, it eventually drops again due to H$_2$ cooling enhanced by X-ray ionization, resulting in the formation of massive BH seed with a mass of $7.6\times10^{4}~M_\odot$.
These results indicate that when the ratio $J_\mathrm{X,21}/J_{21} \gtrsim 10^{-5}$, the thermal evolution deviates 
significantly from the atomic-cooling track ($T \simeq8000~\mr{K}$), preventing the formation of DCBHs.
This trend is consistent with the critical X-ray to UV flux ratio proposed by \citet{Inayoshi_and_Omukai_2011}.
On the other hand, in overdense regions, strong turbulence allows the formation of massive BH seeds with $M_\mathrm{BH} > 10^4~M_\odot$ even under such conditions.
\par
For each combination of baryonic streaming velocity and X-ray intensity, we classify the thermal evolution of collapsing gas in $10^{3}$ merger trees into three categories, following \citet{Li_et_al_2021}: 
(i) the $\chHt$ track, (ii) the $\chH$-$\chHt$ track and (iii) the $\chH$-$\chH$ track.
In cases following the first track (i), $\chHt$ cooling is effective from the early stages, maintaining a low temperature of a few hundred Kelvin.
This behavior is illustrated by the case with $J_\mr{X,21}/J_\mr{21}=10^{-4}$ and $v_\mr{bsm}=0$ in Figure \ref{fig:EvolOnezone}.
In the second track (ii), the gas initially undergoes a short period of atomic cooling before transitioning to $\chHt$ cooling.
This mode is exemplified by the case with $J_\mr{X,21}/J_\mr{21} = 10^{-5}$ and $v_\mr{bsm} = 0$ in Figure \ref{fig:EvolOnezone}.
In the third track (iii), the gas remains at $T\simeq 10^{4}$~K throughout the collapse, corresponding to the atomic-cooling track that can lead to DC.
This track is followed in cases with very weak or no X-ray irradiation, such as $J_\mathrm{X,21}/J_{21} \leq 10^{-6}$ in Figure~\ref{fig:EvolOnezone}.
\par
\begin{figure*}[t]
    \begin{center}
      \includegraphics[width=\linewidth]{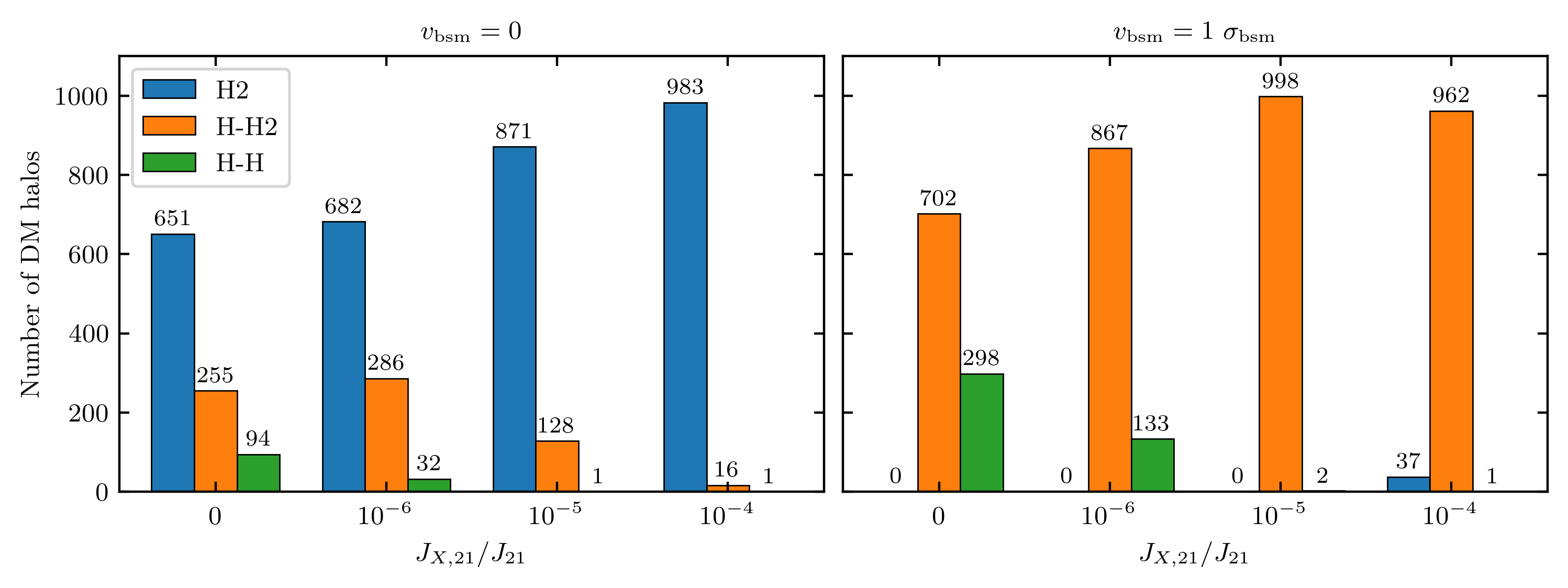}
      \caption{
    The number of DM halos that follow each evolutionary track under different X-ray intensities. 
    All halos evolve into $M_\mathrm{h} = 10^{12}~M_\odot$ at $z = 6$. 
    The left and right panels correspond to the cases with $v_\mathrm{bsm} = 0$ and $1\sigma_\mathrm{bsm}$, respectively. 
    The blue, orange, and green bars indicate the (i) H$_2$ track, (ii) H–H$_2$ track, and (iii) H–H track, respectively. 
    The numbers above the bars represent the number of halos in each category.
    This figure shows how X-ray intensity and baryonic streaming motion influence the thermal evolution pathway of collapsing gas, and thus the likelihood of direct collapse BH formation.
        }
      \label{fig:Classification}
    \end{center}
\end{figure*}

Figure~\ref{fig:Classification} summarizes the number of DM halos that follow each evolutionary track for different X-ray intensities.
The left and right panels correspond to the cases without ($v_\mr{bsm} = 0$) and with baryonic streaming motion ($1\sigma_\mr{bsm}$), respectively.
We first focus on the case without streaming motion (left panel). 
In the absence of X-ray irradiation, 91~\% of the halos experience $\chHt$ cooling 
(i,e, either the H$_2$ or H-H$_2$ track), while only 9.4~\% remain on the H-H track and lead to DC.
As the X-ray intensity increases, the fraction of DC cases drops rapidly.
Even under X-ray to UV flux ratios inferred from local star-forming galaxies ($J_\mathrm{X,21}/J_{21} = 10^{-6}$), only 3.2\% of the halos undergo DC.
For stronger X-ray intensities expected from HERA measurements, $J_\mr{X,21}/J_\mr{21}=10^{-5}$ and $10^{-4}$, 
the DC fraction declines further to a negligible level of $\sim 0.1\%$.
These results indicate that, in the absence of streaming motion, X-ray intensity consistent with HERA observation can strongly suppress the DC channel, rendering it highly inefficient under realistic astrophysical conditions.
\par
In the presence of baryonic streaming motion ($v_\mathrm{bsm} = 1\sigma_\mathrm{bsm}$, right panel), cloud collapse is initially delayed by the relative motion.
In addition, as the halo continues to grow, turbulence driven by halo mergers and irradiation from nearby galaxies becomes stronger. 
These processes prevent the gas from collapsing until the halo reaches the atomic cooling regime ($M_\mathrm{halo} > 10^7~\mathrm{M}_\odot$).
As a result, without X-rays, the H$_2$ track is no longer realized. All halos experience at least one episode of atomic hydrogen cooling and are thus categorized into either the H–H or H–H$_2$ track.
In this case, approximately 30\% of the halos follow the H–H track and can potentially form DCBHs.
However, similar to the no-streaming case, the DC fraction sharply declines with increasing X-ray intensity.
When the X-ray intensity reaches $J_\mathrm{X,21}/J_{21} = 10^{-5}$ or $10^{-4}$, the occurrence of DC drops again to just 0.1\%.
These results clearly demonstrate that, 
even in the presence of streaming motion, which generally favors DCBH formation, the DC channel remains highly sensitive to X-ray irradiation at levels consistent with HERA observations.

\par
\begin{figure*}[t]
    \begin{center}
      \includegraphics[width=\linewidth]{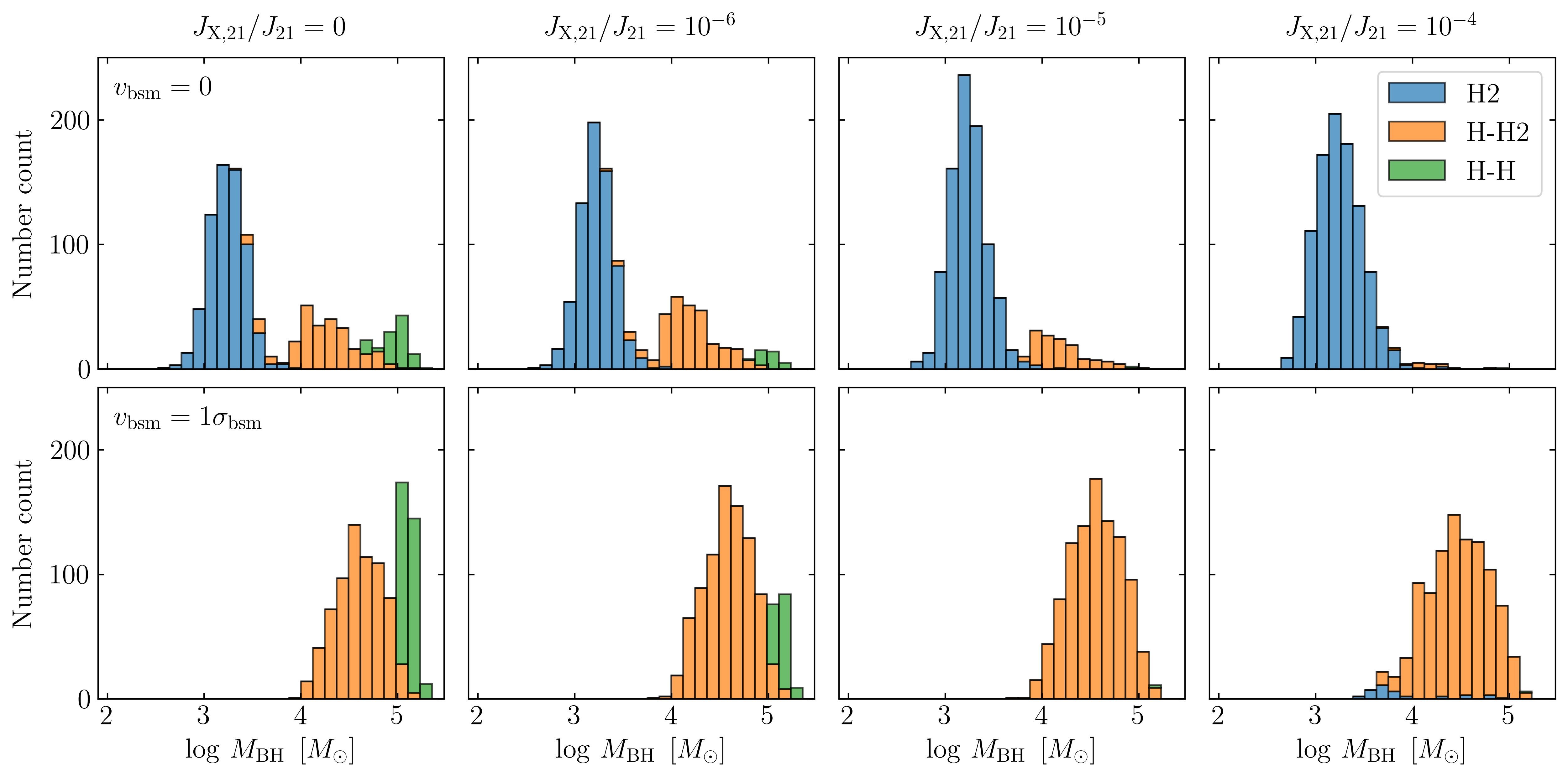}
      \caption{
      Mass distribution of seed black holes formed within DM halos that grow to $10^{12}~M_\odot$ at $z=6$. 
    The contributions from different evolutionary tracks are stacked within each mass bin. 
    The upper and lower panels show the results for $v_\mathrm{bsm} = 0$ and $1\sigma_\mathrm{bsm}$, respectively. 
    From left to right, the X-ray intensity increases: $J_\mathrm{X,21}/J_{21} = 0$, $10^{-6}$, $10^{-5}$, and $10^{-4}$.
    This figure illustrates how baryonic streaming motion and X-ray irradiation affect the mass spectrum of seed black holes by modifying the thermal evolution pathway of the collapsing gas.
      }
      \label{fig:MassDistribution}
    \end{center}
\end{figure*}
\begin{figure*}[t]
    \begin{center}
      \includegraphics[width=\linewidth]{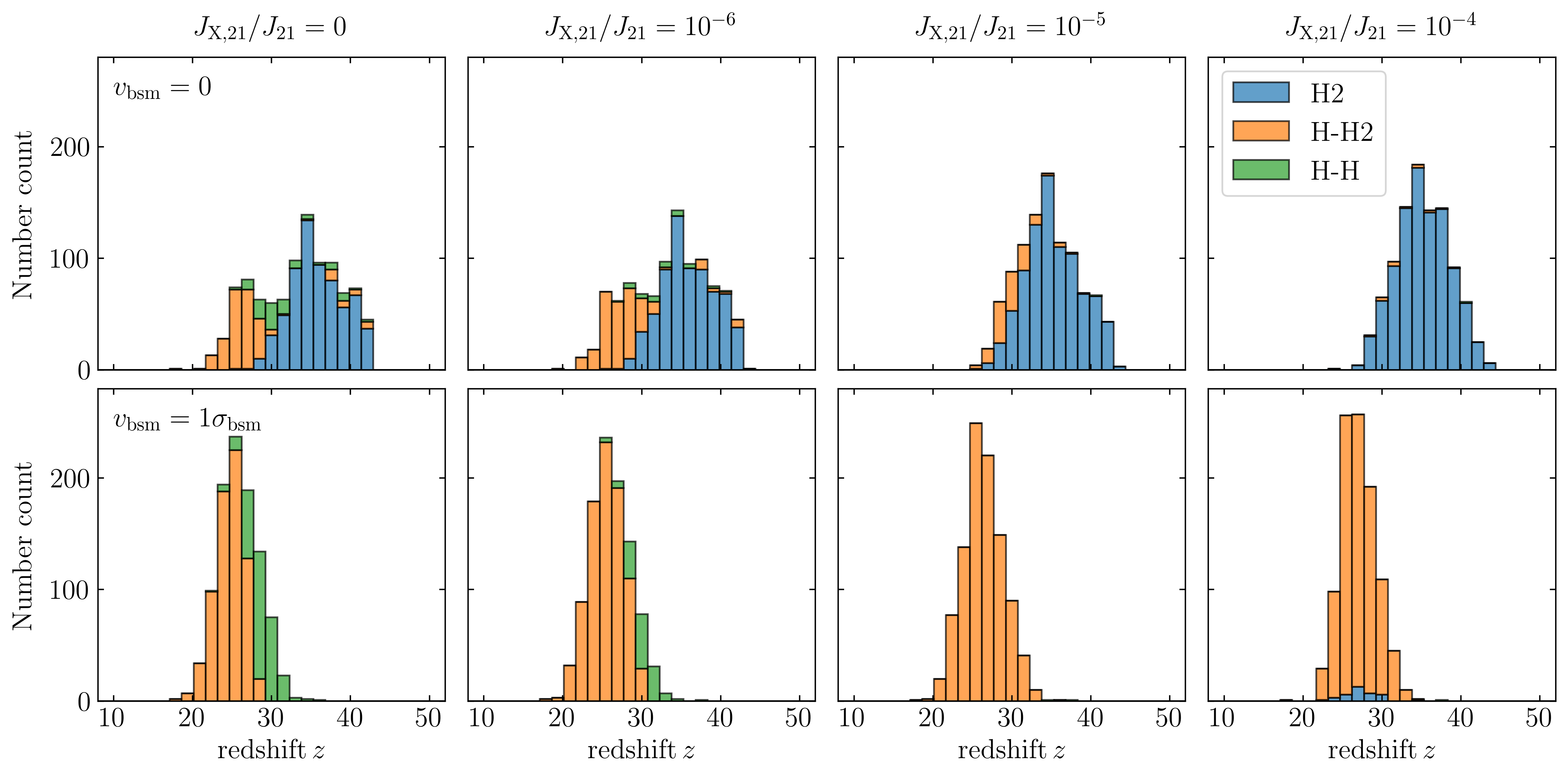}
      \caption{
      Redshift distribution of seed \blackhole formation within DM halos that grow to $10^{12}~M_\odot$ at $z=6$. 
      The panel layout and color scheme are the same as in Figure~\ref{fig:MassDistribution}, but the horizontal axis denotes the formation redshift.
      This figure illustrates how baryonic streaming motion and X-ray irradiation affect the formation redshift by modifying the graviational stability of the gas within DM halos.
      }
      \label{fig:RedshiftDistribution}
    \end{center}
\end{figure*}

Figure \ref{fig:MassDistribution} shows the mass distribution of seed \blackholes under various combinations of baryonic streaming velocities and X-ray intensities.
The top and bottom panels show the results without and with streaming motion.
The contributions from different evolutionary tracks are stacked within each mass bin.
The top-left panel shows the results without streaming motion and X-ray. 
In this case, each of the three evolutionary tracks contributes to a distinct peak, and the \blackhole mass shows a broad distribution from 
$4.1\times10^2~M_\odot$ to $1.7\times10^5~M_\odot$.
As shown in the top panels, increasing the X-ray intensity suppresses the occurrence of $\chH$-$\chHt$ and $\chH$-$\chH$ tracks, thereby
reducing the formation frequency of massive \blackhole seeds with $M_\mathrm{BH} > 10^4~M_\odot$.
Under the X-ray strength suggested by HERA observations, $J_\mr{X,21}/J_{21}\geq10^{-5}$, the $\chHt$ track is dominant. 
As a result, the mass distribution exhibits a prominent peak around $10^3~M_\odot$, and massive seeds with $M_\mr{BH}>10^{4}~M_\odot$ 
are rarely formed.
\par
The bottom left panels of Figure \ref{fig:MassDistribution} show the results for the case with $v_\mr{bsm}=1\sigma_\mr{bsm}$ 
and no X-ray irradiation.
As indicated in the right panel of Figure~\ref{fig:Classification}, the presence of baryonic streaming motion enhances the frequency of the $\chH$–$\chHt$ and $\chH$–$\chH$ tracks.
This results in a double-peaked mass distribution in the high-mass regime where $M_\mr{BH}>10^{4}~M_\odot$.
With increasing X-ray intensity, the number of $\chH$-$\chH$ tracks decreases. 
Nevertheless, unlike the case with $\sigma_\mr{bsm}=0$, the majority of merger trees still follow the H–H$_2$ track.
Consequently, although the number of \blackholes with masses exceeding $10^5~M_\odot$ becomes less frequent, 
the mass distribution retains its peak around $3 \times 10^4~M_\odot$ as shown in the bottom panels.
This indicates that although strong X-ray irradiation suppresses DC formation channels, $1\sigma_\mr{bsm}$ streaming motion can still commonly produce massive BH seeds with $M_\mathrm{BH} \gtrsim 10^4~M_\odot$.
\par
Figure~\ref{fig:RedshiftDistribution} shows the redshift distribution of seed BH formation. 
While the panel layout and color scheme are the same as in Figure~\ref{fig:MassDistribution}, the horizontal axis here represents the formation redshift.
As shown in the top-left panel, seed formation along the $\chHt$ tracks predominantly occurs at early epochs ($z \sim 30$--$45$), while the $\chH$--$\chHt$ tracks appear more frequently at later times ($z \sim 20$--$30$). The $\chH$--$\chH$ tracks are most common at intermediate redshifts around $z \sim 30$.
This trend arises because the radiation intensity from nearby galaxies increases with time at $z \gtrsim 30$, reaches a peak, and then gradually declines (see Figure~2 in \citeauthor{Li_et_al_2021} \citeyear{Li_et_al_2021}).
When halos become gravitationally unstable before the radiation intensity builds up, they tend to follow the $\chHt$ track under weaker radiation backgrounds.
On the other hand, halos that collapse when the radiation field is near its peak tend to follow the $\chH$–$\chH$ track.
\par
As shown in the top panels, increasing the X-ray intensity eventually leads to nearly all halos experiencing the $\chHt$ track at $z\gtrsim30$.
This is because X-ray irradiation promotes $\chHt$ formation, enabling the gas within halos to become gravitationally unstable earlier through $\chHt$ cooling.
In contrast, the bottom panels show that, for the case with $v_\mathrm{bsm} = 1\sigma_\mathrm{bsm}$, nearly all halos form BHs at later times ($z < 30$), regardless of the X-ray intensity.
This is because turbulence induced by the baryon streaming motion prevents the gas from becoming gravitationally unstable at earlier epochs.
Additional turbulence from mergers and enhanced radiation from nearby galaxies develop during this period and further delay the collapse.
As a result, collapse occurs only at later times, when the radiation background is strong enough to suppress the $\chHt$ track.
\subsection{Cosmological abundance of BH seeds} \label{sec:BHMF}
In this section, we calculate the cosmic abundance of \blackhole seeds in a comoving volume.
Our calculations provide the distribution of BH seed masses and their formation redshifts as a function of host halo mass and baryonic streaming velocity.
By combining these results with halo abundance statistics, we can construct the mass function of BH seeds.
As discussed in Section~\ref{subsec:SeedBHForm}, we consider DM halos that grow up to $M_\mr{h}=10^{11}$, $10^{12}$, and $10^{13}~M_\odot$ at $z=6$. 
We estimate their number densities using the Sheth-Tormen halo mass function \citep{Sheth_et_al_2001} as
\begin{eqnarray}
    n_{M_\mr{h}} = \int_{M_\mr{h}}^{10M_\mr{h}} \f{dn_\mr{ST}}{dM_\mr{h}^\prime}~dM_\mr{h}^\prime .
\end{eqnarray}
The resulting comoving number densities are $9.9\times10^{-4}$, $6.1\times10^{-6}$, and $8.4\times10^{-10}~\mr{Mpc}^{-3}$, respectively.
For each halo mass, we consider two representative cases with $v_\mathrm{bsm} = 0$ and $1\sigma_\mathrm{bsm}$.
To obtain the total BH seed abundance, we assume that 60~\% of halos have no streaming velocity and the remaining 40~\% experience a velocity of $1\sigma_\mathrm{bsm}$, motivated by the fact that approximately 60~\% of the cosmological volume has streaming velocities below $1\sigma_\mathrm{bsm}$ \citep{Schauer_et_al_2021}.
This approach provides a conservative estimate of the impact of streaming motion.
Note that in this work we do not include halos with masses below $10^{11}~M_\odot$. These low-mass halos are much more numerous and significantly contribute to the overall BHMF. Nevertheless, since quasars observed at $z > 6$ reside in halos with masses above $10^{11}~M_\odot$ \citep{Shimasaku_and_Izumi_2019}, we focus on this mass range in the present study.
\par
\begin{figure*}[t]
  \begin{center}
    \includegraphics[width=0.8\linewidth]{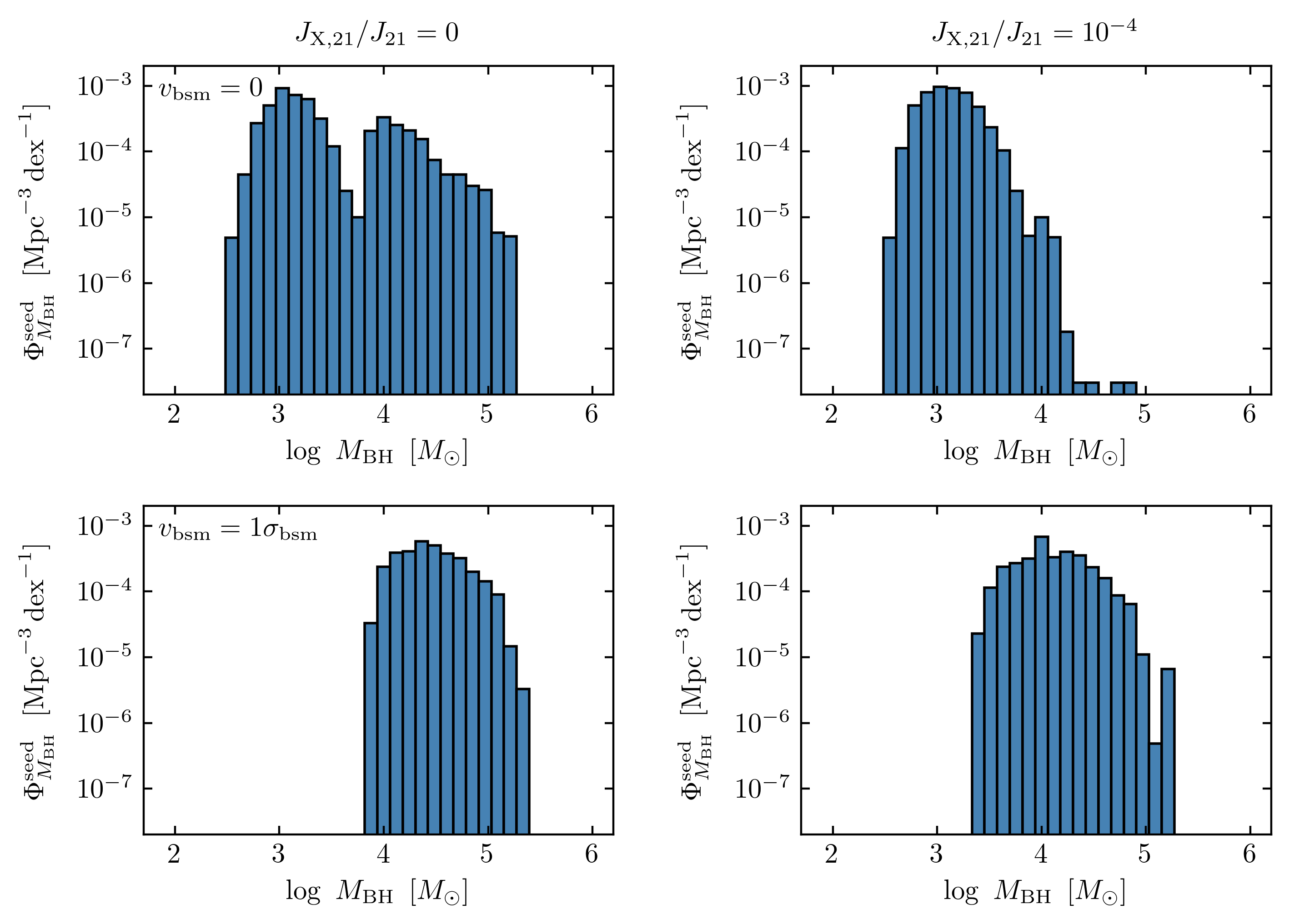}
    \caption{
        Seed BH mass functions (BHMFs) at $z > 15$ for different parameter combinations. 
    The top and bottom rows correspond to the cases with $v_\mathrm{bsm} = 0$ and $1\sigma_\mathrm{bsm}$, respectively. 
    The left and right columns represent the cases with $J_\mathrm{X,21}/J_\mathrm{21} = 0$ and $10^{-4}$, respectively. 
    The combined BHMFs for $v_\mathrm{bsm} = 0$ and $1\sigma_\mathrm{bsm}$ are shown in Figure~5.
    This figure highlights how baryonic streaming motion and X-ray intensity independently and jointly influence the shape and amplitude of the seed BHMF at high redshift.
    }
    \label{fig:NumberDensity}
  \end{center}
\end{figure*}

Figure~\ref{fig:NumberDensity} shows the seed BHMFs $\Phi_{M_\mr{BH}}^\mr{seed}$ for different baryonic streaming velocities and X-ray intensities. 
In our model, BHs form over a wide redshift range of $z = 15\text{--}45$, but we show the mass function combining all those BHs, without accounting for their subsequent growth.
As seen in the top-left panel, in the case of $v_\mr{bsm} = 0$ and $J_\mr{X,21}/J_{21} = 0$, the mass function exhibits a bimodal distribution.
The lower-mass peak originates from the $\chHt$ track, where efficient H$_2$ cooling leads to low-temperature evolution and the formation of low-mass \blackholes.
In contrast, the high-mass peak is associated with the $\chH$–$\chHt$ and $\chH$–$\chH$ tracks, where atomic cooling leads to higher gas temperatures and the formation of more massive seeds.
When the X-ray intensity is strong ($J_\mathrm{X,21}/J_{21} = 10^{-4}$), as shown in the top-right panel, the high-mass peak disappears.
This is because X-ray irradiation enhances the production of $\chHt$ and thus lowers the gas temperature.
As a result, the formation of massive \blackholes above $10^4~M_\odot$ is significantly suppressed.
\par
In contrast, baryonic streaming motion regulates the BHMF shape, producing a single peak even under strong X-ray irradiation.
As shown in the bottom-left panel, in the case with  $v_\mr{bsm} = 1\sigma_\mr{bsm}$ and $J_\mr{X,21}/J_{21} = 0$, 
the BHMF exhibits a single peak around $3 \times 10^4~M_\odot$.
This is because most merger trees follow either the H–H$_2$ or H–H track due to the delay of collapse caused by streaming motion.
Even when the X-ray intensity increases, the distribution remains single-peaked since the H–H$_2$ track continues to dominate.
However, the enhanced H$_2$ cooling efficiency slightly lowers the gas temperature, causing the peak to shift modestly toward lower masses.
Thus, baryonic streaming motion allows the BHMF to extend up to $\sim 10^5~M_\odot$ even in the presence of strong X-ray irradiation.

\section{DISCUSSIONS} \label{sec:discussion}
\subsection{Comparison with observed BHMF}
Thanks to its unprecedented sensitivity, JWST has detected less massive BHs in the high-redshift universe than previously achievable. 
With the growing number of such detections, it has recently become feasible to place constraints on their mass function. 
Notably, at $z \sim 4$, the BHMF has been constrained down to BH masses of $\sim 10^6~M_\odot$ \citep{Matthee_et_al_2024,Taylor_et_al_2024}.
These observations have motivated a number of theoretical studies aiming to reproduce the observed BHMF and investigate the formation and growth pathways of high-redshift SMBHs \citep{Li_et_al_2024,Liu_et_al_2024,Jeon_et_al_2025}.
Here, we compare the BHMFs predicted by our model with the observational one.
Although our model covers only lower-mass BHs at higher redshifts and is therefore not directly comparable to the observed BHMF, it provides the initial conditions for the subsequent growth of BHs that may eventually be observed.
\par
\begin{figure*}[t]
  \begin{center}
    \includegraphics[width=\linewidth]{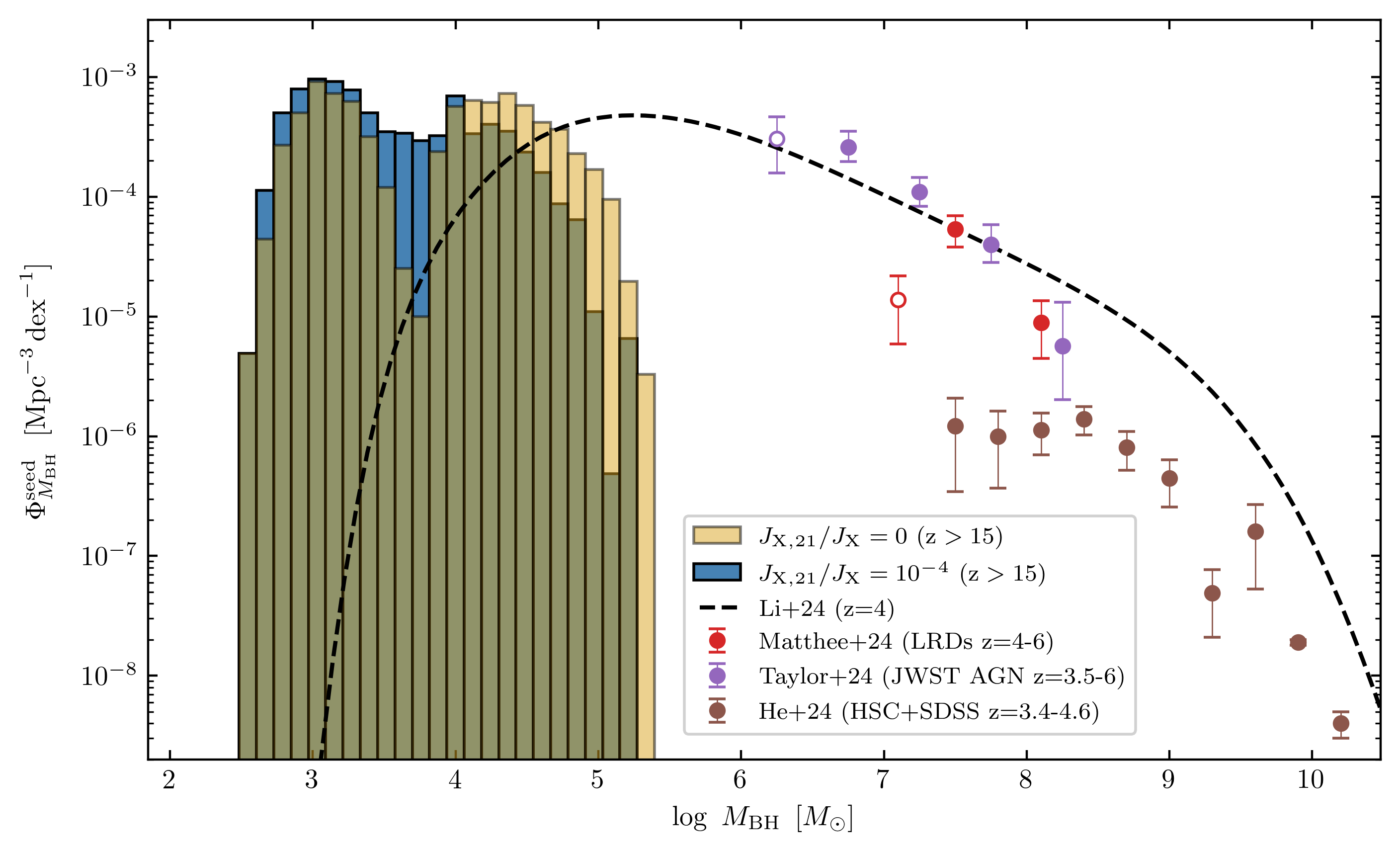}
    \caption{
    Seed BH mass functions (BHMFs) at $z > 15$ for the cases with $J_\mathrm{X,21}/J_\mathrm{21} = 0$ (beige bars) and $10^{-4}$ (blue bars), calculated for halos that grow to $M_\mathrm{h} = 10^{11}$--$10^{13}~M_\odot$ at $z=6$.
    The dashed line shows the BHMF at $z = 4$ predicted by the theoretical model of \citet{Li_et_al_2024}. 
    Red, purple, and brown points represent the observational BHMFs at $z = 3\text{--}6$ from Little Red Dots \citep{Matthee_et_al_2024}, JWST broad-line AGNs \citep{Taylor_et_al_2024}, and HSC+SDSS quasars \citep{He_et_al_2024_Quasar}, respectively. 
    Open symbols (purple and red) indicate data points that are affected by sample incompleteness.
    This figure compares our seed BHMF predictions with observations of overmassive black holes at lower redshifts, illustrating that early-formed seeds can account for a significant fraction of the observed population if efficiently grown.
    }
    \label{fig:NumberDensity_obs}
  \end{center}
\end{figure*}

In Figure \ref{fig:NumberDensity_obs}, we show our total BHMFs (beige and blue bars), obtained by summing up the BHMFs for the $v_\mr{bsm} = 0$ and $1\sigma_\mr{bsm}$ shown in Figure~\ref{fig:NumberDensity}, along with observational data at $z\sim3\text{--}6$ (colored points).
As can be seen by comparing the blue and beige bars, the presence of X-rays reduces the number of massive BHs with $M_\mr{BH}\gtrsim10^4~M_\odot$.
However, even in the case with X-rays, integrating our BHMF above $10^4~M_\odot$ yields a number density of $\sim10^{-4}~\mr{Mpc}^{-3}$.
On the other hand, the observed number density of BHs with masses above $10^{6}~M_\odot$ is also estimated to be $\sim10^{-4}~\mr{Mpc}^{-3}$.
This suggests that, if the massive BH seeds formed in our model at $z >15$ grow efficiently thereafter, they could account for the observed population of SMBHs with $M_\mr{BH}>10^{6}~M_\odot$  at later times $z\sim3\text{--}6$.
In particular, since our BHs are formed in overdense regions, they are expected to grow more efficiently and are likely to contribute to the high-mass end of the observed BHMF.
In principle, seed BHs with masses below $10^4M_\odot$ could grow into the observed SMBHs if they accrete efficiently.
However, such growth is generally constrained not only by the Eddington accretion limit, but also by low duty cycles of accretion episodes over extended periods, due to feedback processes that can regulate or interrupt gas inflow.
Even though earlier formation of light seeds is taken into account (as shown in Figure\ref{fig:RedshiftDistribution}), massive seed formation remains a more favorable pathway for explaining the presence of SMBHs at high redshifts.
For these reasons, our study focuses on the conditions where seed BHs with $M_\mathrm{BH} \gtrsim 10^4~M_\odot$ can still form even in the presence of strong X-ray backgrounds.
\par
The black dashed line in Figure \ref{fig:NumberDensity_obs} shows the model prediction from \citet{Li_et_al_2024}.
This curve represents the predicted BHMF resulting from the growth of BH seeds originally proposed in \citet{Li_et_al_2023}.
As shown in the figure, their model can successfully reproduce the observed BHMF, suggesting that the seed formation and growth scenarios in such overdense region may account for the high-redshift SMBH population.
The evolution and observability of lower-mass seed black holes formed under strong X-ray backgrounds, however, remain open questions for future work.

\subsection{Overmassive BH formation}
Recent JWST observations have revealed an increasing number of  active galactic nuclei (AGNs) at high redshifts with measurements of both BH mass
and galaxy host stellar mass. 
The majority of these AGNs have been reported to host overmassive BHs with BH-to-stellar mass ratios exceeding the empirical values observed in the local universe,
$M_\mr{BH}/M_*\sim 10^{-3}$ \citep[e.g.,][]{Harikane_et_al_2023_AGN,Pacucci_et_al_2023,Maiolino_et_al_2024}.
Although it remains debated whether this apparent overmassive nature is intrinsic or influenced by observational biases, 
the existence of BHs more massive than expected from the empirical mass ratio is now confirmed.
Several scenarios have been proposed to explain the origin of these BHs, including super-Eddington accretion, suppressed star formation 
in their host galaxies, and the initially massive BH seeds \citep{Inayoshi_et_al_2022_SMBH,Hu_et_al_2022,Scoggins_et_al_2024,Pacucci_et_al_2024}.

\par
To compare our theoretical model with observed overmassive SMBHs, we estimate the stellar mass of each host galaxy at the BH formation epoch in our model.
Our model provides the BH mass, its formation redshift, and the growth history of the host halos.
Assuming the cosmic baryon fraction of $f_\mathrm{b} = 0.157$ and a star formation efficiency (SFE) of $\epsilon_* = 0.1$, 
the stellar mass of the host galaxy is estimated as $M_* = \epsilon_* f_\mathrm{b} M_\mathrm{h}$.
While typical SFE values inferred from local observations are only a few percent, significantly higher values up to $\sim 0.3$–$0.5$ have been reported in extreme environments such as starbursts and super star-clusters \citep{Murray_et_al_2010}.
In addition, such values are required to reproduce the UV luminosity functions of high-redshift bright galaxies observed by JWST \citep{Harikane_et_al_2023_galaxy,Inayoshi_et_al_2022_SF}.
Moreover, recent numerical simulations suggest that such high efficiencies are achievable in the early universe, where galaxies are expected to have high gas surface densities and low metallicities \citep{Fukushima_and_Yajima_2021,Chon2024,Menon_et_al_2024}.
Based on these considerations, we adopt $\epsilon_* = 0.1$ as our fiducial value.
If the actual SFE is lower, the stellar mass of the host galaxies would be smaller, further enhancing the overmassive nature of 
the BHs relative to their hosts.

\par
\begin{figure*}[t]
  \begin{center}
    \includegraphics[width=\linewidth]{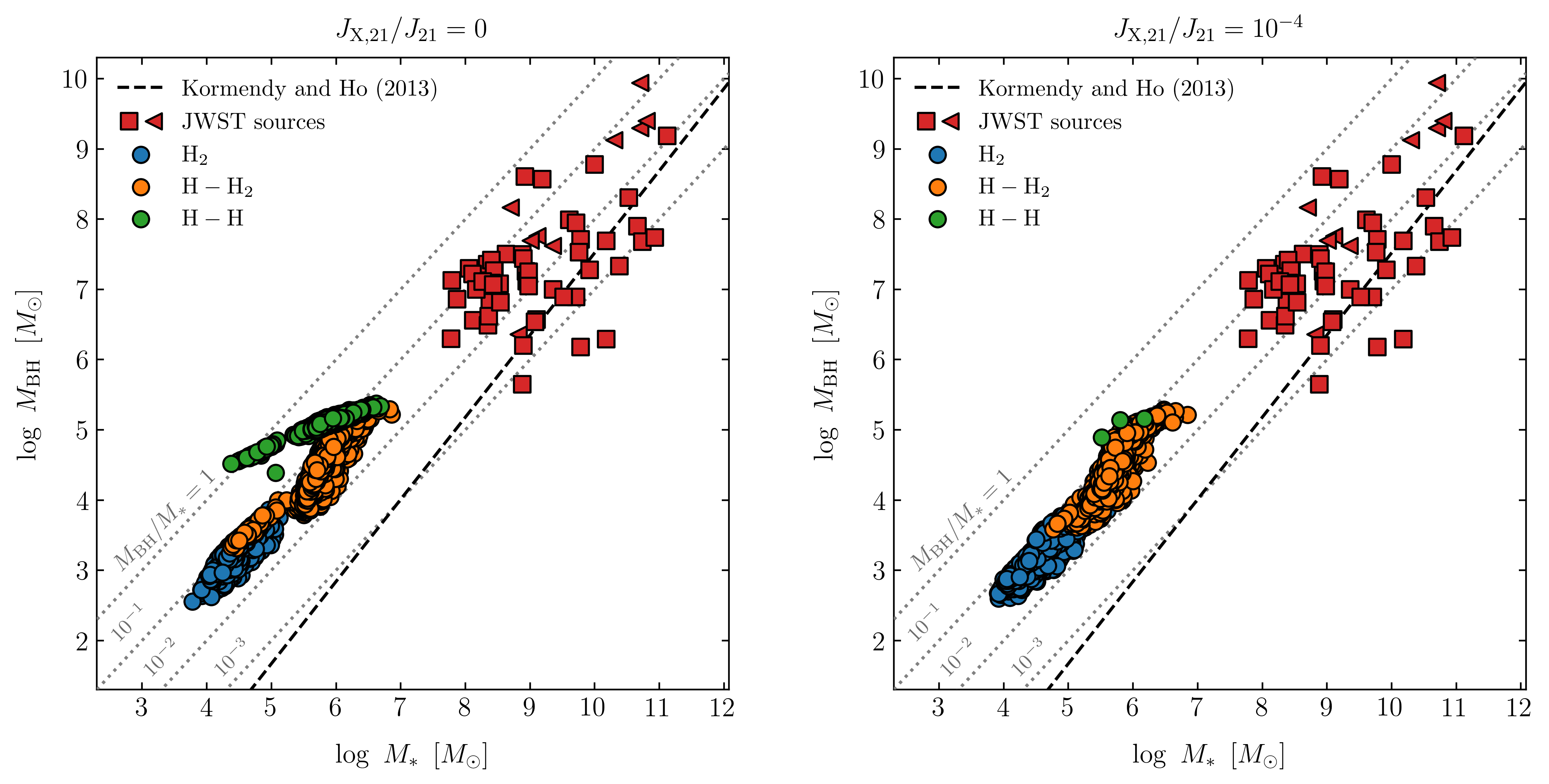}
    \caption{
        Relation between BH mass and total stellar mass of host galaxies ($M_\mathrm{BH}$--$M_*$) at the time of BH formation. 
    The left and right panels correspond to the cases with $J_\mathrm{X,21}/J_\mathrm{21} = 0$ and $10^{-4}$, respectively. 
    The colored circles show our model results, with colors indicating the H$_2$ (blue), H–H$_2$ (orange), and H–H (green) tracks. 
    The total stellar mass of the host galaxy is estimated as $M_* = \epsilon_* f_\mathrm{b} M_\mathrm{h}$, assuming $f_\mathrm{b} = 0.157$ and $\epsilon_* = 0.1$.
    Red points represent SMBHs observed by JWST \citep{Kokorev_et_al_2023, Harikane_et_al_2023_AGN, Ding_et_al_2023, Stone_et_al_2023, Stone_et_al_2024, Maiolino_et_al_2024, Juodzbalis_et_al_2024_nature, Juodzbalis_et_al_2025}. 
    Squares indicate sources with both $M_\mathrm{BH}$ and $M_*$ measurements, while left-pointing triangles show upper limits on $M_*$. 
    The dashed line denotes the local $M_\mathrm{BH}$--$M_\mathrm{*,bulge}$ relation from \citet{Kormendy_and_Ho_2013}, and the gray dotted lines indicate constant $M_\mathrm{BH}/M_*$ ratios.
    This figure shows that our model can naturally produce black holes exceeding the local $M_\mathrm{BH}$--$M_*$ relation, consistent with overmassive SMBHs observed at high redshift.
        }
    \label{fig:Mbh_Mstar}
  \end{center}
\end{figure*}

Figure \ref{fig:Mbh_Mstar} shows the $M_\mr{BH}-M_*$ relation based on our BH samples (colored circles), along with the JWST observational data \citep[red squares and triangles,][]{Kokorev_et_al_2023,Harikane_et_al_2023_AGN,Ding_et_al_2023,Stone_et_al_2023,Maiolino_et_al_2024,Juodzbalis_et_al_2024_nature,Juodzbalis_et_al_2025} and lines of constant mass ratio $M_\mr{BH}/M_*$ (diagonal dotted lines).
In our sample, BHs formed via the $\chHt$, $\chH$–$\chHt$, and $\chH$–$\chH$ tracks are shown in blue, orange, and green, respectively.
The red squares indicate JWST sources with both $M_\mr{BH}$ and $M_*$ estimates, while the left-pointing triangles represent those with upper limits on $M_*$.
The left and right panels show the cases without X-ray irradiation and with the strongest X-ray intensity in our models
($J_{\rm X,21}/J_{21}=10^{-4}$), respectively.
Most BHs in our sample exhibit mass ratios in the range of $M_\mathrm{BH}/M_* \sim 10^{-2}-10^{-1}$, nearly an order 
of magnitude higher than the local empirical relation \citep[dashed line;][]{Kormendy_and_Ho_2013}.
In the absence of X-ray irradiation (left panel), \blackholes formed through the $\chHt$ and $\chH$–$\chHt$ tracks lie 
$\sim 2$~dex above the local $M_\mr{BH}–M_*$ relation, while those formed via the $\chH$-$\chH$ track exhibit even larger offsets.
In the presence of strong X-ray irradiation (right panel), the $\chH$–$\chH$ track is significantly suppressed, 
limiting the formation of extremely massive BHs with $M_\mathrm{BH}/M_* \simeq 1.0$. 
Nonetheless, even in this case, the BH seeds remain overmassive, with typical offsets of $\sim 2$~dex.
If the actual SFE at high redshift is closer to the local value of a few percent, the stellar masses would be lower, causing the BHs to appear even more overmassive.
\par
Our results suggest that, even in the presence of strong X-ray irradiation in the early universe, seed \blackholes that lie above the local $M_\mr{BH}-M_*$ relation can naturally form in the overdense regions we have focused on.
These \blackholes can explain a large fraction of the overmassive systems observed by JWST even if the $M_\mr{BH}/M_*$ ratio slightly decreases over time, 
and can account for the most extreme outliers if they maintain the ratio established at birth.
Therefore, within the context of the overdense regions of the universe considered in this study, additional mechanisms that would further increase 
the $M_\mr{BH}$ to $M_*$ ratio, such as super-Eddington accretion or suppressed star formation, are not necessary to account for the observed population of overmassive BHs.

\subsection{The uncertainty in the effect of baryonic streaming motion} \label{subsec:bsm}
Baryonic streaming motion has a significant impact on gas dynamics within halos.
In our model, this effect is incorporated through the parameter $\alpha_0$ (see Equation~\ref{eq:c_eff}). 
We adopt a fiducial value of $\alpha_0 = 4.7$ in this work. However, recent cosmological hydrodynamic simulations suggest that its value lies in the range from 1 to 10 \citep{Hirano_et_al_2017, Schauer_et_al_2019, Hirano_et_al_2025}.
Given this uncertainty, it is important to examine how our results change depending on the assumed value of $\alpha_0$.
\par
If a smaller value of $\alpha_0$ is adopted, the delay of gas collapse is less significant. 
As a result, the cloud collapses in halos at an earlier cosmic time, when the LW radiation is still relatively weak. 
This leads to an increased number of lower-mass BHs formed via the H$_2$ track, with masses in the range of $10^3-10^4~M_\odot$.
However, the emergence of overmassive BHs in our model does not depend on whether H$_2$ cooling is effective. Therefore, even in this case, our conclusion that overmassive BHs can naturally form in overdense regions remains unchanged, although the contribution from the $\chHt$ track becomes more significant.
\par
On the other hand, if a larger value of $\alpha_0$ is adopted, the collapse time is further delayed.
If the delay were too large, the LW intensity would decrease due to the dilution effect of cosmic expansion.
However, since the delay is not significant even with $\alpha_0$ increasing up to $\sim10$, the LW intensity does not change substantially.
As a result, the thermal evolution of the collapsing gas remains nearly unchanged, and the overall outcome of BH seed formation is not substantially affected.

\section{Summary} \label{sec:summary}
We have investigated the formation of black hole (BH) seeds in overdense regions in the early universe by extending the semi-analytic model developed by \citet{Li_et_al_2021} to include the effects of X-ray irradiation. Motivated by recent constraints from Hydrogen Epoch of Reionization Array (HERA), which imply that X-ray intensities in the early universe may be significantly higher than previously expected, we examine how enhanced X-ray irradiation affects the thermal evolution of collapsing gas within halos and the resultant BH seed masses.
We explore a range of X-ray intensities that spans values inferred from both local observations and HERA constraints, and calculate the resulting BH mass function.

\par
We found that X-ray irradiation at the intensity levels inferred from recent HERA observations significantly alters the thermal evolution and formation pathways of BH seeds.
In the absence of streaming motion, X-ray ionization enhances $\chHt$ formation, leading the gas to follow the $\chHt$ cooling track, similar to that of typical Population~III star formation and resulting in low-mass BH seeds.
In the presence of streaming motion, although the direct collapse (DC) channel is suppressed under strong X-ray irradiation, atomic $\chH$ cooling becomes effective during the early stages of collapse, followed by $\chHt$ cooling at higher densities.
This hybrid thermal evolution enables the formation of relatively massive BH seeds with masses larger than $10^{4}~M_\odot$, even in X-ray-rich environments.
\par
Despite the suppression of DC owing to X-ray irradiation, our model demonstrates that massive BH seeds can still form in significant numbers.
In particular, the comoving number density of BH seeds with masses above $10^4~M_\odot$ is comparable to that of the observed supermassive BHs with $M_\mr{BH}$ at $z \sim3\text{--}6$ ($\sim 10^{-4}~\mr{Mpc}^{-3}$).
Furthermore, we find that BH seeds can have $M_\mr{BH}$-$M_*$ ratios of $\sim0.01\text{--}0.1$ at their birth, placing them significantly above the local $M_\mr{BH}$–$M_*$ relation. 
These results suggest that they are promising progenitors of the supermassive BHs recently reported by JWST.

\begin{acknowledgments}
We thank Wenxiu Li and Shingo Hirano for valuable discussions, insightful comments, and for providing their code and data.
This work was supported by JSPS KAKENHI Grant Numbers JP24KJ0015 (KK) and JP22H00149 (KO).
We also acknowledge support from the National Natural Science Foundation of China (Grant No. 12233001), the National Key R\&D Program of China (Grant No. 2022YFF0503401), and the China Manned Space Project (CMS-CSST-2025-A09).
\end{acknowledgments}

\bibliography{bib}{}
\bibliographystyle{aasjournal}



\end{document}